# Reliable Multicast in Large Enterprise Wireless Networks

Abhijeet Bhorkar and Gautam Bhanage
1322 Crossman Ave, Sunnyvale, CA
Aruba Networks*

## Motivation:

Multicast applications such as live stream of a football match, IP-based radio stations, webcast of live sessions are ever increasing. In recent years there has been significant commercial interest in using multicast traffic from public-service broadcasters and other DSL TV providers and are expected to grow.

This puts pressure on the last-mile wireless technologies such as wireless enterprise network in utilizing the bandwidth efficiently and delivering the multi-cast traffic in reliable manner. Furthermore, wireless multicast transmissions are unreliable as they do not provide any feedback from receiver and thus the onus is on the transmitter for reliable transmission.

We first discuss with the issues in current approaches used in delivering multicast traffic in enterprise wireless networks:

a) **Unreliable transmission of multi-cast traffic:**
   Current approaches for reliable multicast communications in enterprise wireless networks includes choosing reliable multi-cast transmission rate. This option of choosing multi-cast transmission rate is still unreliable as there is no acknowledgement from the clients.
b) **Inefficient use of bandwidth:** To increase the reliability of the multi-cast traffic, existing techniques convert multi-cast flow into unicast flows for each client. This option of using unicast traffic to all the clients is not a scalable solution. Furthermore, converting the multicast traffic into the unicast traffic we perform the retransmission of the same data linearly to the number of clients.

## Contributions:

- We propose a new architecture Rate-less Code based Multicast (RCNC) for of transmission of multi-cast traffic in an efficient and reliable manner in enterprise wireless network. The underlying design makes use of an upcoming application layer rate-less codes such as raptor codes[1].

- The performance gains for the rate-less codes have been demonstrated theoretically by the research community. However, there are many challenges still left to successfully deploy network coding based solutions in real world applications. This work tries to bridge the gap between employing network [1]coding in practice and real world applications.

- Current wireless client do not have support to utilize the coding mechanism. Here, we propose a method to enable the clients to decode the rate-less codes.
- We also determine the conditions when the rate-less codes should be utilized rather the conventional methods. In this way, we propose a fall-back mechanism back to conventional mechanism in case the computation complexity of rate-less decoders is higher than the most clients can support it.
- The architecture is has less control overhead.

## Rate-less codes:

Rate-less coding received a tremendous response in improving capacity of wireless networks. 3rd Generation Partnership Project (3GPP) uses the rate-less codes (Raptor code) in its specification for Multimedia Broadcast/Multicast Services (MBMS) [2], the Digital Video Broadcasting project (DVB), which uses the Raptor code in the specification for handheld devices (DVB-H) [1] and for IPTV [3].

**Useful properties of Rate-less codes:**
Consider a simple example of transmission of multicast traffic to N clients from AP. For the multicast traffic is converted into unicast, east packet is retransmitted N times. Furthermore, there is huge explosion of ACK and retransmission for a single packet. However, with rate-less codes the data is encoded intelligently and only one single ACK is required from each client for the entire chunk of transmitted data. This reduces significant overhead of traffic without any comprise on reliability.

**Rate-less codes in operation:**
Rate-less codes encode large chunk data into small blocks. The transmitter transmits random linear combinations of these smaller blocks until the receivers receive all chunks. A receiver is able to recover the file successfully when receiving distinct encoded packets. These codes are shown to be information theoretically optimal and capacity achieving. Rate-less codes have desirable property of rate-less-ness, i.e. they have theoretically the ability to generate an unlimited amount of uniquely encoded data on-the-fly.

Let us consider an example of rate-less codes (this example shows non-optimal implementation).
A large block of data X is split into k segments $X_1, X_2, \ldots X_k$, modulated with say BPSK (finite field of GF(2)). These segments are then encoded into m > k messages $\{Y_1, Y_2, \ldots Y_m\}$ as the following random linear combinations $Y_i = \sum_i^k = \beta_i X_i$, where $\beta_{i,j}$ are randomly chosen elements in the finite field over GF(2). The parameters $\beta_{i,j}$ of the encoding are adjusted so that the rows $[\beta_{i,1}, \beta_{i,2}, \ldots \beta_{i,k}]$ are linearly independent with high probability. Thus, any host that receives k of the $Y_i$'s can solve the corresponding system of linear equations to determine X. If the m encodings $Y_1 \ldots Y_m$ prove insufficient (due to poor channel conditions, for example), then the encoding node can easily generate a number of extra packets $Y_i$ by using newly constructed random elements $\beta_{l,j}$.

Value of number of segments k depends on the delay requirements at the application. E.g., for the video traffic encoded at say 10Mbps, and packet sizes of 1.5k bytes, we need to encode about 84 packets each second assuming the maximum latency allowed at 1 seconds.

Figure 2 graphically demonstrates the encoding process.

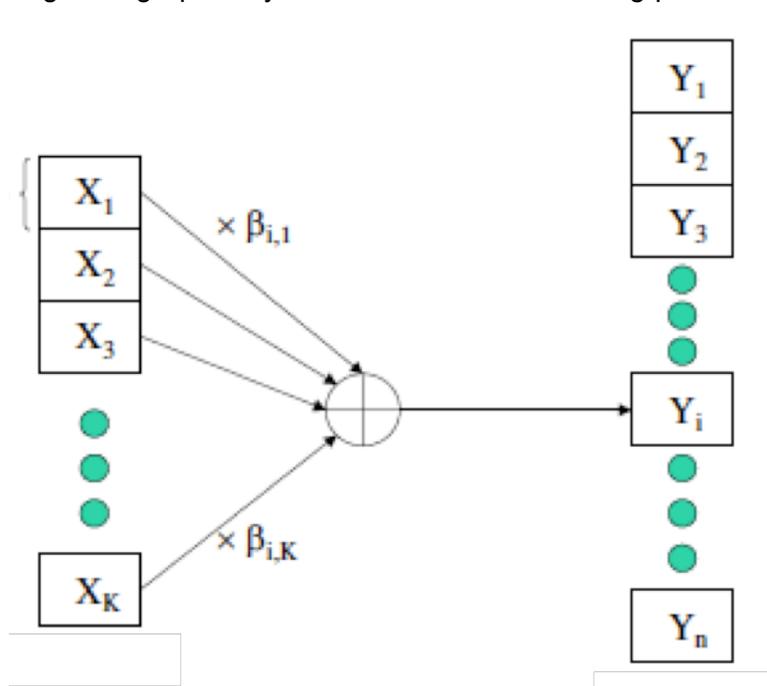

At the receiver, once the node has received k encoded packets it attempts to decode using matrix inversion and Gaussian elimination. Once the decode is successful the receiver delivers the data to the multicast application. Otherwise the node discards any linearly independent packets and waits for more encoded packets.

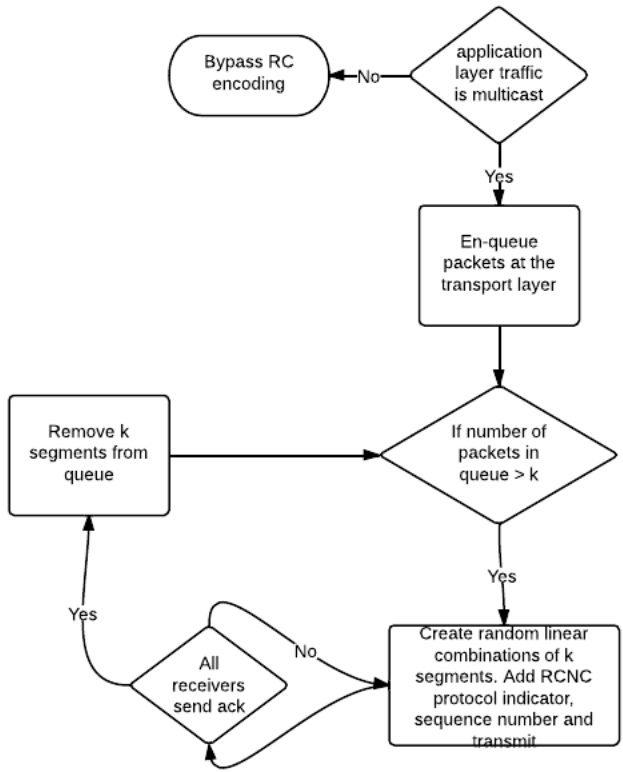
Sender Design

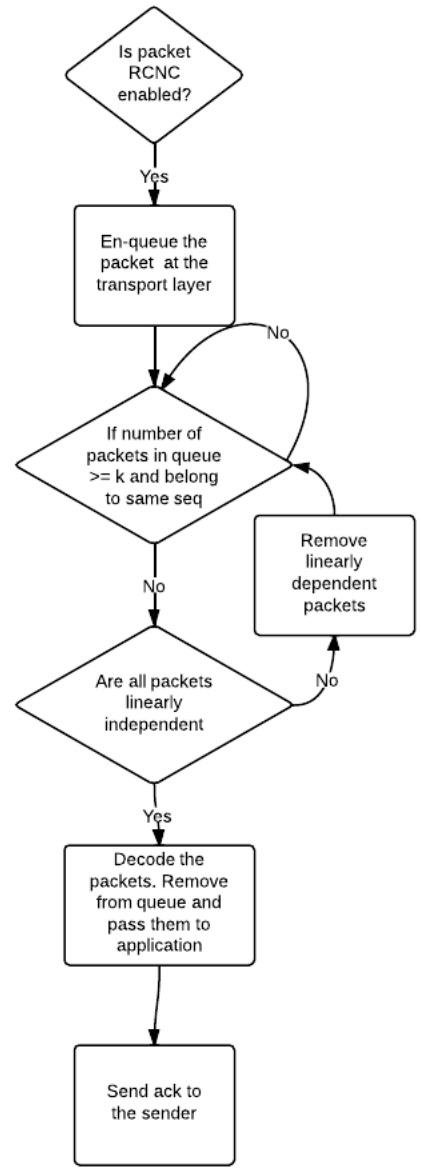
Receiver Design

In this work, we will be studying the problem of transmitting multicast packets to large number of clients connected to an access point in an enterprise wireless networks. Note that, this work can further be extended to the situations where network coding.

## Design:

In this section, we will consider design and implementation issues for implementation the RCNC protocol.
1) Clients do not have a mechanism for decoding rate-less codes
   Clients may have less computational power to decode the packets encoded with rate-less codes.
2) Control overhead for enforcing the Rate-less codes may be higher than conventional methods. In this case, the AP has to decide if it wishes to continue with the using rateless codes.

**Pushing the rate-less decoder into the clients:**
   When the client connects to the AP, the network management device pushes the decoder into the client. Note that client may accept to download the decoder based on the agreement between the network manager. If client accepts to negotiate the capabilities, during the agreement between the client negotiates the capabilities with the network manager. Network manager finally decides if the client has enough resources to push the decoder into the client. The exact mechanism for pushing the decoder is beyond the scope of this. A mechanism to push the decoder application describe in Figure 3.

**Decision to employ rate-less codes**
- Use of rateless codes is most beneficial where the number of multi-cast clients is huge. Typically, 30-40 clients is a decent number. If the number of multi-cast client is less than a particular threshold (say 10), using the conventional method of unicast is good comprise for the gains vs. computation complexity involved.
- Rate-less codes may not utilized if the client does not support the decoding ability. In this case, we must use unicast packets for the weak clients and use RCNC for the clients which support RCNC decoding.
- RCNC works on the principle that the associated clients receive the packets independently. This means that if the clients are collocated, it might be a good idea to only employ unicast packets for reliable communications.

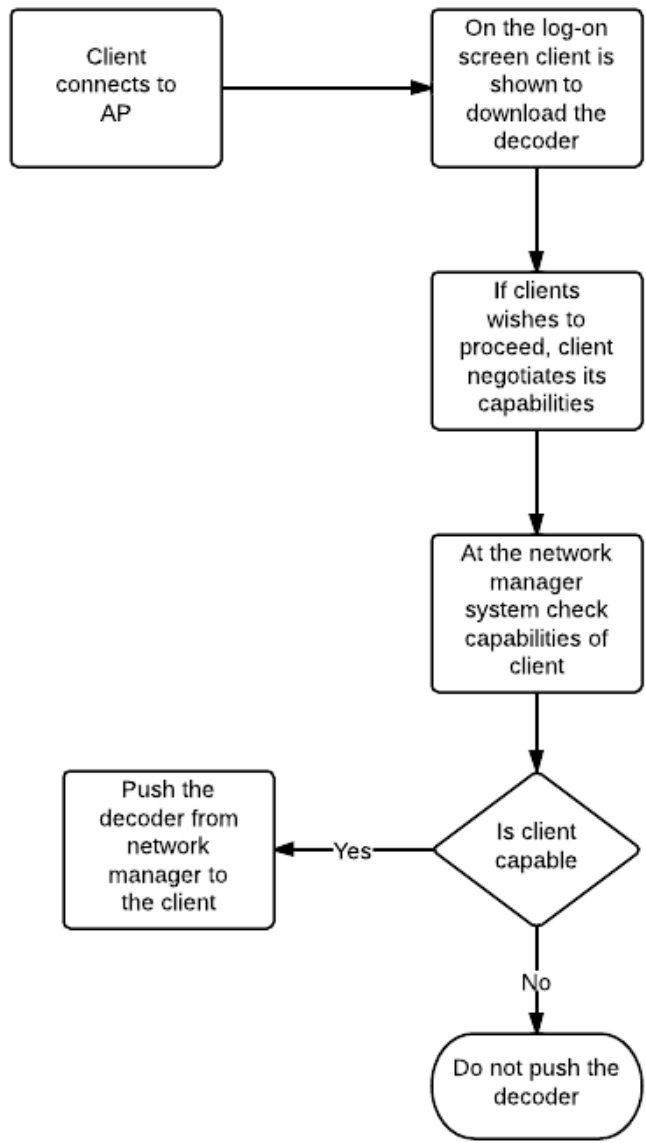

**Figure 3**

Figure 4. shows the overall mechanism and architecture to employ RCNC.

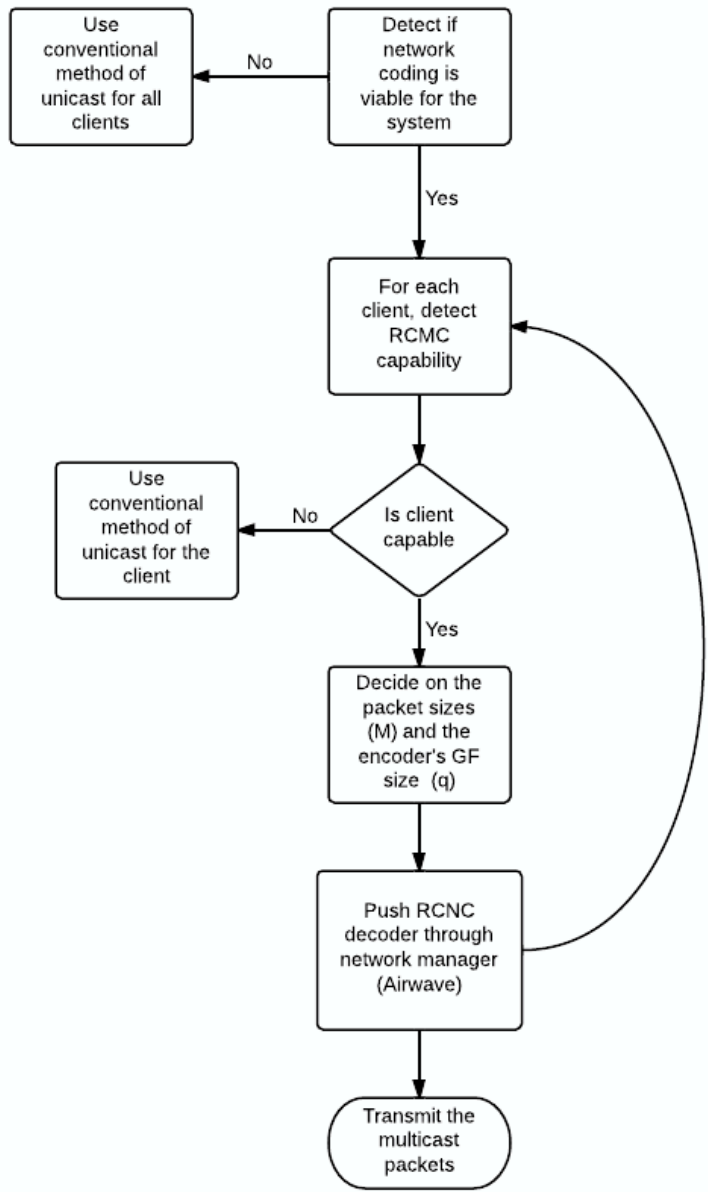

**Fig 4.**

## Performance:

We consider the performance of a simple network where N clients are connected to an access point.

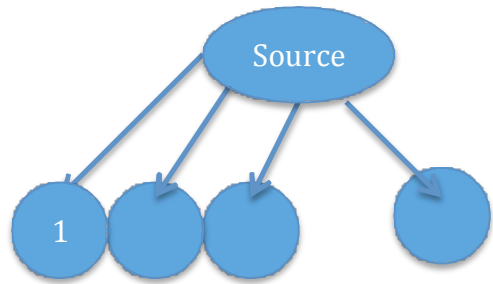

Fig .5

We consider point-multipoint multicast transmission of UDP traffic. We assume that the each packet transmitted is successful with probability p. We plot the overhead of the multicast transmissions over the RCNC mechanism in Fig. 6 when p=1/2, in terms of total air-time incurred for the transmission. We assume that for the unicast packets the Access points undergoes exponential back off.

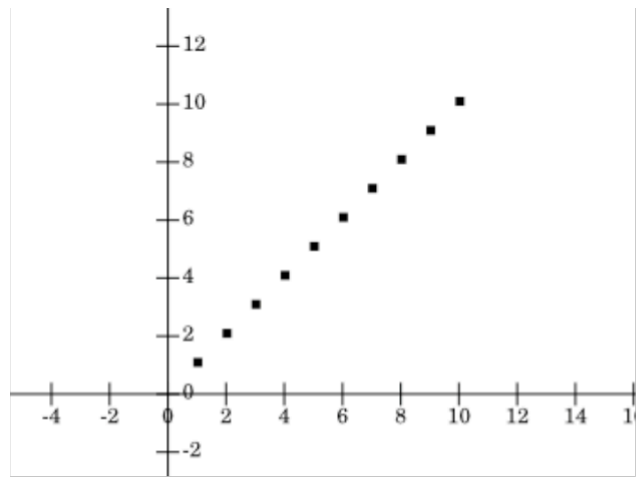

## Conclusions:

In this work, we have presented a complete end-end architecture to employ the rate-less codes. We have developed a new architecture Rate-less Codes Multi-cast (RCNC).
This architecture is shown to provide high throughput gains, reliability and near optimal throughput performance.

We have used a simple version of rate-less codes mostly adapted from rate-less deluge[4].
This version of rate-less codes is less efficient. However, newer codes such Raptor-codes[3] can be used at the cost of increased encoder-decoder efficiency.